# AHA-GRAPE: Adaptive Hydrodynamic Architecture – GRAvity PipE

T. Kuberka, A. Kugel, R. Männer, H. Singpiel
Dept. for Computer Science V,
University of Mannheim,
B6-26, 68131 Mannheim, Germany

R. Spurzem
Astronomisches Rechneninstitut, Mönchhof-
strasse 12-14,
69120 Heidelberg, Germany

Ralf Klessen
Sterrewacht Leiden, Postbus 9513,
2300 RA Leiden, The Netherlands

**Abstract** In astrophysics numerical star cluster simulations and hydrodynamical methods like SPH require computational performance in the petaflop range. The GRAPE[1]-family of ASIC-based accelerators improves the cost-performance ratio compared to general purpose parallel computers, however with limited flexibility.

The AHA-GRAPE architecture adds an reconfigurable FPGA[2]-processor to accelerate the SPH computation. The basic equations of the algorithm consist of three parts each scaling with the order of $O(N)$, $O(N*Nn)$ and $O(N^2)$ respectively, where N is in the range of $10^4$ to $10^7$ and $Nn \sim 50$. These equations can profitably be distributed across a host workstation, an FPGA processor and a GRAPE-subsystem.

With the new ATLANTIS FPGA-processor we expect a scalable SPH-performance of 1.5Gflops per board. The first prototype AHA-GRAPE system will be available in mid 2000. This 3-layered system will deliver an increase in performance by a factor of 10 as compared to a pure GRAPE solution.

*Keywords:* Stellar Dynamics, Hydrodynamics, Star Clusters, Numerical Methods, FPGA

## 1 Introduction

### 1.1 Multi-particle interactions

The gravitating N-body-problem is one of the grand challenges of theoretical physics and astrophysics. Its accurate solution for very large particle numbers cannot generally be obtained by mathematical considerations (series evaluations) as it was possible for the historical treatment of the classical two- and three-body problems. Only computer modeling on the fastest available hardware using specialized mathematical-numerical algorithms can be used as an appropriate tool here.

Hydrodynamical problems fall into the same category: Analytic solutions exist only for a limited number of highly simplified cases. Hence, understanding the time evolution of a gaseous system in most cases involves sophisticated numerical modeling.

The questions related with astrophysical system typically involve addressing both sets of problems simultaneously. For instance, clusters of stars form from collapse and fragmentation of self-gravitating gas clouds and grow in mass by accretion of the available gas reservoir. In their later dynamical evolution stellar clusters dissolve due to a combination of close and distant encounters between the stars. Eventually they blend into the overall stellar distribution of the Milky Way.

In addition to the fundamental theoretical interest of large gravitating N-body systems such models are essential for our understanding of the structure and evolution of many astrophysically relevant objects, as there are our planetary system, our own and other galaxies and the entire universe seen as an object forming structure via gravitational interaction between particles. Such numerical modeling is also important for the interpretation of a wealth of new observational data from space based instruments, as e.g. of the dense centers of galactic nuclei observed with the Hubble Space Telescope (HST). They have recently improved evidence for the existence of supermassive black holes in their centers [26]. Surrounding them is a very dense star cluster, rotating, axisymmetric if not triaxial. Despite recent attempts to tackle this problem the physical interplay between relaxation, star accretion and black hole growth in such situations remains an unsolved and challenging theoretical and numerical problem for the astrophysical N-body simulators.

---

[1] GRAPE: **Gra**vity **P**ip**e**: An ASIC for parallel calculation of the gravitational force [25].
[2] FPGA: **F**ield **P**rogrammable **G**ate **A**rray. FPGAs are the core elements of reconfigurable custom computing machines.

For this physical situation a particular class of `high-accuracy' numerical models following the orbit of each particle due to the `exact' gravitational forces of all the other particles in a many-body system has to be used. We use a fourth order Hermite predictor-corrector scheme with hierarchically blocked individual time steps, Ahmad-Cohen (AC) neighbor scheme, and regularization of close encounters [27] and hierarchical subsystems [28][29]. This method is widely known as Aarseth scheme [1][2][33]. Furthermore, particle based methods can also be applied to solve the equations of hydrodynamics. A widely used scheme is SPH (smoothed particle hydrodynamics). In this approach the fluid is represented by an ensemble of particles each carrying mass and momentum (analog to the N-body problem) and additional properties like temperature, pressure, entropy, and so forth. Thermodynamic observables are determined in a local averaging process over a given set of neighboring particles [6][30]. This method is fully Lagrangian and is successfully applied in various fields of numerical astrophysics. It is especially useful when dealing with the interaction between stellar and self-gravitating gaseous systems, as it elegantly unites the hydrodynamical and the gravitational N-body approach within one numerical scheme.

## 1.2 Implementations

The algorithms require most of their computational time to accumulate the mutual pairwise gravitational forces between the particles (N² problem!) and to compute a list of neighbors. Until recently this limited the maximum particle number for SPH calculation to ~$10^5$ even on large super computers [36]. The constraint is even more severe in the case of collision dominated N-body calculations. For example, three years ago the record particle number used to follow a globular cluster into core collapse was only $10^4$ particles [30]. This situation improved considerably with the advent of the special purpose computers of the GRAPE series which were developed in Japan [25] and are also used in Germany and many other countries in the world. In the context of globular cluster simulations this pushed the record to 32k particles and proved the existence of gravothermal oscillations [22]. Despite being constructed to solve the gravitational N-body problem with high speed, GRAPE devices are useful for a large variety of other astrophysical applications as well, ranging from cosmological problems [5][30] down to studies of the dynamical friction of a binary black hole in galactic nuclei [24][23]. In particular, the combination with the particle based SPH method has opened the door to also study hydrodynamical problems with GRAPE. For example, it has been used to investigate the properties of X-ray halos around galaxy clusters [1], or the properties of interstellar turbulence [5][30]. GRAPE has proven to be especially useful for hydrodynamical collapse calculation in the context of star and planet formation [17][5].

However, the special purpose machines of the GRAPE series reach their highest efficiency only for problems, which can be tackled with pure and clean N-body algorithms such as N-BODY4 or KIRA. For SPH or standard N-body simulations using an AC neighbor scheme or a very large number of close (so-called primordial) binaries, or even worse for molecular dynamics simulations with potentials other than the Coulomb potential (e.g. van der Waals) they are not the optimal choice.

One commonly used solution is to use general purpose massively parallel machines as the CRAY T3E, for which a competitive implementation of N-BODY6++ exists using MPI and SHMEM [2]. While its performance compares well with one of the single GRAPE-4 boards, a larger scale GRAPE machine or the coming GRAPE-6 are still much more efficient for the pure N-body case. There is still work in progress, however, to improve the implementation on the general purpose parallel computers.

The new solution presented here is to build a hybrid machine, which uses for the intermediate range forces a reconfigurable custom computing machine: an FPGA processor. This new system will profit from both the extremely high performance of the GRAPEs for the $O(N^2)$ gravitational force computation and the high degree of flexibility of the FPGA processor which lets it adapt to the needs of the various hydrodynamic (SPH) oriented computations in the $O(N*Nn)$ region.

### 1.3 FPGA processors

The family of FPGA devices was introduced in 1984 by Xilinx. FPGAs feature a large number of relatively simple elements with configurable interconnects and an indefinite number of reconfiguration cycles with short configuration times. All configuration information is stored in SRAM cells. The basic processing element[3]

- Increased density by a factor of 24 from 1993 through 1998 (Xilinx XC4000: 400 to 18400 elements)
- increased speed by a factor of 3 from 1994 through 1998 (Xilinx XC4000: 133 to 400MHz internal toggle rate).

Eight years of experience at the University of Mannheim with FPGA based computing machines shows that this new class of computers is an ideal concept for constructing special-purpose processors combining both the speed of a hardware and the flexibility of a software solution [19][19]. The so called FPGA processors consist of a matrix of FPGAs and memory forming the computational core. In addition there are a (programmable) I/O unit and an internal (configurable) bus system. As processing unit, I/O unit and bus system are implemented in separate modules, this kind of system provides scalability in computing power as well as I/O bandwidth.

FPGA processors have shown to provide su-

Table 1: FPGA Processor Specs

| | |
|---|---|
| Number of FPGAs per computing board | 4 |
| Memory size per computing board | 40MB |
| Computing board memory bandwidth | 4 GB/s @ 50MHz |
| Max. number of FPGA boards per PCI bus | 7 |
| Max. PCI bandwidth | 125MB/s @ 33MHz, 32Bit PCI |
| PCI bandwidth @4k blocks | 75MB/s @ 33MHz, 32Bit PCI |
| Private bus bandwidth | 800MB/s @ 50MHz per board pair |
| Expected SPH floating-point performance per computing board | 1.5Gflops @ 50MHz |
| I/O board external bandwidth | 4*200MB/s @ 50MHz |
| Number of supported host CPUs per PCI | 1 |
| Supported OS | Win NT, Linux |

(PE) of all current mainstream FPGAs is a 4-input/1-output look-up-table (LUT) with an optional output register. The functionality of the FPGA is thus determined by the contents of the look-up-tables within the PE's and the "wiring" between these elements.

Over the last few years FPGA performance has increased tremendously as it profits from both:

---
[3] The currently largest devices offer approx. 10.000 of these PEs and more than 400 I/O pins.

perior performance in a broad range of fields, like encryption, DNA sequencing, image processing, rapid prototyping etc. Very good surveys can be found in [9] and [7]. The hybrid microprocessor/FPGA systems developed at the University of Mannheim are in particular suitable for:

- acceleration of computing intensive pattern recognition tasks in High Energy Physics (HEP) and Heavy Ion Physics,
- subsystems for high-speed and high-frequency I/O in HEP,

- 2-dimensional industrial image processing,
- 3-dimensional medical image visualization [11] and
- acceleration of multi-particle interaction calculations in astronomy.

A well-tried means to adjust a hybrid system to different applications is modularity. ATLANTIS implements modularity on different levels. First of all there are the main entities host CPU and FPGA processor which allow to partition an application into modules tailored for either target. Next the architecture of the

## 2 Architecture

### 2.1 FPGA performance assessment

Using FPGAs to accelerate complex computations using floating-point algorithms has not been considered a promising enterprise in the past few years. The reason is that general floating-point as well as particular N-Body implementation have shown only poor performance[4] on FPGAs. Usually N-Body calculations and particle based hydrodynamical simulations

Figure 1: Implementaion of SPH-Fragment on the FPGA-Processor Enable++

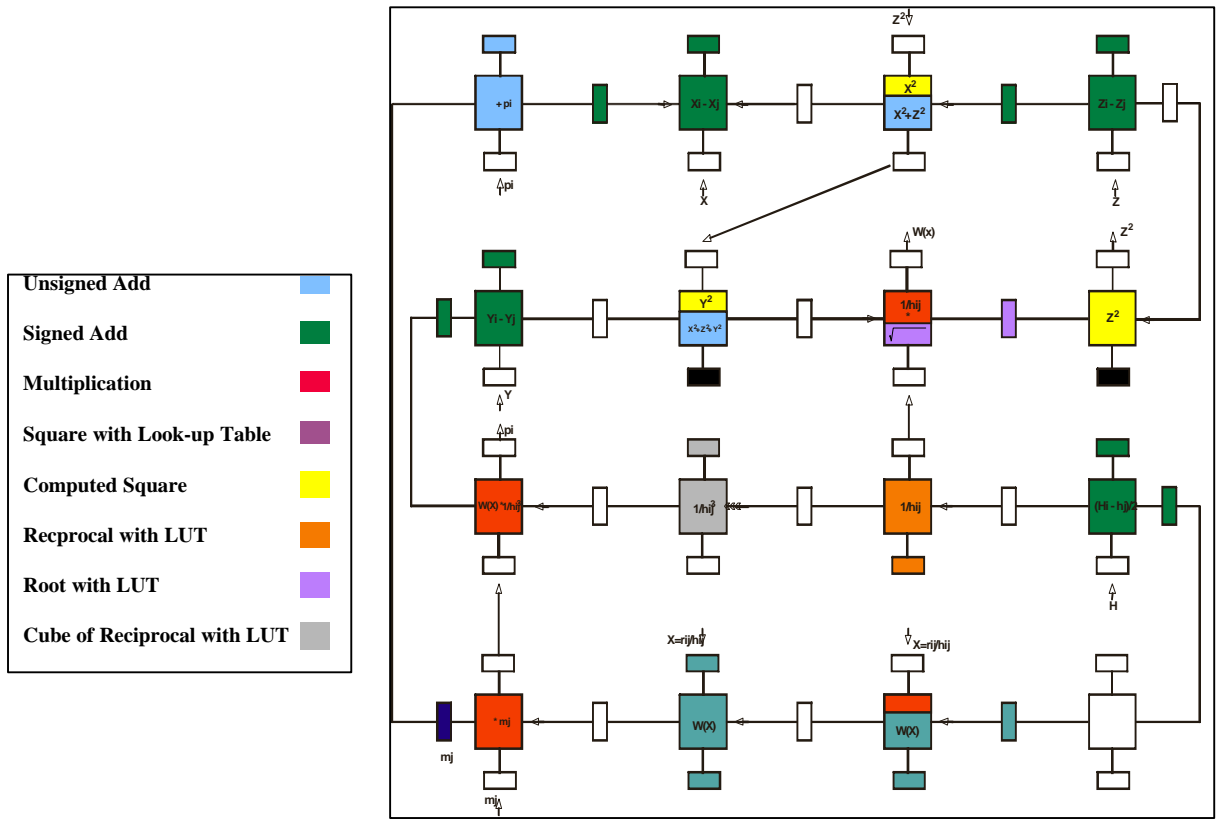

FPGA processor uses one board-type to implement mainly computing tasks and another board-type to implement mainly I/O oriented tasks. A backplane based interconnect system provides scalability and supports an arbitrary mix of the two board-types. Finally modularity is used on the sub-board level by allowing different memory types or different I/O interfaces per board type. The most important parameters of the system are listed in Table 1.

need a computing performance in at least teraflop range and are accelerated with the help of ASIC-based co-processors like the GRAPE-series. Nonetheless we have recently investigated the performance of a certain sub-task of the SPH algorithm on the Enable++ system [12]. The results indicate that FPGAs can indeed provide even in this area a significant per-

---
[4] In 1995 approx. 10 Mflops [31] per Xilinx chip were reported for 18 bit precision, and 40 Mflops [14] with 32 bit precision on an 8 chip Altera board.

formance increase. The piece of code shown in Figure 1 was implemented on 15 out of 16 core FPGAs[5] of the Enable++ system making heavy use of the configurable interconnect structure, as shown in Figure 1. For the implementation a 28bit floating-point format was used: 1 sign-bit, 7 bits exponent, 20 bits mantissa. The maximum pipeline depth is 6 stages and a result is produced at every clock cycle. The total performance for the code in the loop is therefore 16*13MHz = 208Mflops with the XC4013-5

Figure 2: SPH Code Fragment

```
Do i = 1 , N
  Do j = 1 , Nn
    r_ij = r_i - r_j   /*(3-d vectors)*/
    r_ij = | r_ij |
    h_ij = (h_i+h_j)/2
    1/h_ij
    W(r_ij, h_ij)   /*(table look-up)*/
    r_i = r_i + m_j W(r_ij, h_ij)/ h_ij^3
  Enddo
Enddo
```

chips and 16*32MHz = 512Mflops with the XC4028-2 chip respectively. If the XC4036-3 implementation will allow – as we expect – that 2 instances can run in parallel, the performance will increase to 1.024 Gflops. Parallel I/O is also done with 52 or 128MB/s on the input side plus a few MB/s on the output side. ATLANTIS will support two instances of this code to run in parallel on one computing board.

## 2.2 AHA-GRAPE

For astrophysical particle simulations including self-gravity, the determination of the gravitational potential at each particles position is usually the most expensive step in terms of computational time required. This step shall be done by the special hardware GRAPE for force computation in N-body simulations, which proved highly efficient in the case of a pure point-mass simple algorithm (N-BODY1) case. For many more realistic applications however, some parts of the code become important bottlenecks if the gravitational force calculation is done very fast. They are usually of order $O(N*N_n)$ – where $N_n$ is a neighbor particle number and $N_n \ll N$ – and comprise

1. Computation of the neighbor force when using a more complicated, but more efficient N-body algorithm (about 20 flops per pairwise force, of which order $N_n$ per particle per time-step have to be computed).
2. Computation of the kernel function, its derivatives, and terms related to gas dynamical quantities[6] in the SPH algorithm (about 100 flops per pairwise particle interaction, of which again $O(N_n)$ per particle per time-step have to be computed[7]).
3. Integration of binary motions in regularized coordinates as a function of near perturbers (order $N_n$); this is a very sophisticated algorithm and cannot easily be estimated now in its complexity [27].
4. Integration of SCF force (self-consistent field [9] to compute approximately the gravitational potential of distant particles ) for hybrid N-body models[8].

The floating point operations related to 1) and 2) are in principle straightforward to map onto an FPGA processor, however critical for the performance is the word length which is sufficient for each component of the sum of pairwise forces and other SPH expressions. Test calculations have to be performed to clarify this.

The two subsystems – GRAPE cluster and FPGA processor – will be connected to the host workstation by the PCI bus, either directly or via an interface. Within the subsystems the respective local buses will be used to broadcast sample data and intermediate results. In a second step and by close cooperation with the Tokyo group a hierarchical coupling of the FPGA device with GRAPE, including memory and control of the GRAPE, could be envisaged. This will further improve performance by parallelization of force computations and data communication. Figure 3 displays the performance estimates for various systems with and without FPGA processor.

---

[5] The present system uses Xilinx XC4013 FPGAs. A new system is currently being assembled equipped with XC4036 FPGAs.

[6] Like density, pressure, viscosity, energy fluxes, etc., with summation over $N_n$.

[7] Calculation of 1) can be done as a subset of operations within 2) for a combined high-precision gravity SPH-code.

[8] The complexity of this algorithm differs from the standard scalings.

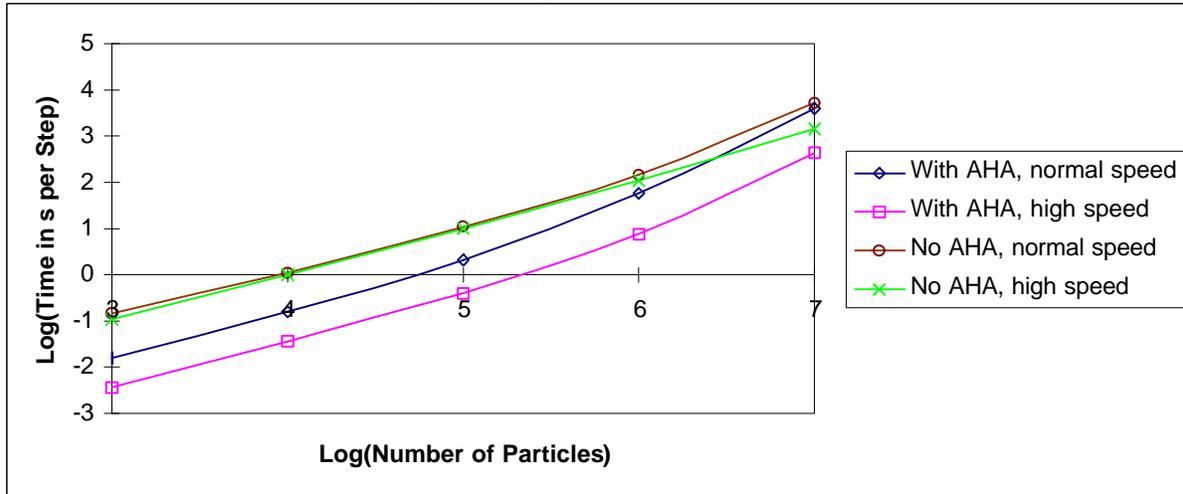

Figure 3: Expected SPH-Performance

## 3 Status and plans

At present (February 1999) a test implementation of the SPH-loop/step1 on ENABLE++ is carried out to verify the estimated performance. By mid 99 the new ATLANTIS system will be available where the full SPH-code has to be implemented. A communication library for LINUX must be developed, supporting simultaneous transfers between host/GRAPE and host/FPGA respectively. We expect the first prototype AHA-GRAPE system to be available in mid 2000. The key figures for this prototype are 50Mflops for the host workstation, 5Gflops for the FPGA processor and 500Gflops for the GRAPE subsystem. The presence of the FPGA processor will lead to an increase in performance by a factor of 10 and will allow us to handle up to approx. $10^6$ particles[9] in collision dominated N-body simulations and a few $10^7$ particles in SPH.

---

[9] At very large particle numbers the $N^2$ term becomes dominant which is the domain of the GRAPE subsystem